# Decoding Human and AI Persuasion in National College Debate: Analyzing Prepared Arguments Through Aristotle's Rhetorical Principles

Mengqian Wu*[1], Jiayi Zhang[2], Raymond Z. Zhang[3]

[1]McGill University
[2]University of Pennsylvania
[3]Stanford University

mengqian.wu@mail.mcgill.ca

**Abstract.** Debate has been widely adopted as a strategy to enhance critical thinking skills in English Language Arts (ELA). One important skill in debate is forming effective argumentation, which requires debaters to select supportive evidence from literature and construct compelling claims. However, the training of this skill largely depends on human coaching, which is labor-intensive and difficult to scale. To better support students in preparing for debates, this study explores the potential of leveraging artificial intelligence to generate effective arguments. Specifically, we prompted GPT-4 to create an evidence card and compared it to those produced by human debaters. The evidence cards outline the arguments students will present and how those arguments will be delivered, including components such as literature-based evidence quotations, summaries of core ideas, verbatim reading scripts, and tags (i.e., titles of the arguments). We compared the quality of the arguments in the evidence cards created by GPT and student debaters using Aristotle's rhetorical principles: ethos (credibility), pathos (emotional appeal), and logos (logical reasoning). Through a systematic qualitative and quantitative analysis, grounded in the rhetorical principles, we identify the strengths and limitations of human and GPT in debate reasoning, outlining areas where AI's focus and justifications align with or diverge from human reasoning. Our findings contribute to the evolving role of AI-assisted learning interventions, offering insights into how student debaters can develop strategies that enhance their argumentation and reasoning skills.

## 1. Introduction

Debate is a powerful tool in English Language Arts (ELA) for developing students' critical thinking, argumentation, and rhetorical skills. However, selecting effective arguments typically relies on one-on-one or small-group coaching, which is labor-intensive and hard to scale. Large language models (LLMs) like GPT offer a scalable alternative, but their ability to generate arguments on par with human debaters remains uncertain. This study investigates GPT's capacity to select compelling evidence and craft high-quality arguments, focusing on its use in generating evidence cards during debate preparation. We employed both quantitative and qualitative methods to evaluate GPT's rhetorical effectiveness. Qualitatively, we developed a codebook grounded in Aristotle's classic rhetorical appeals, including ethos, pathos, and logos (Braet, 1992). Quantitatively, we used indicators derived from prior research (Moon et al., 2024) to identify patterns in student-produced content reflecting these rhetorical strategies. Our integrated framework combines thematic analysis with measurable subdimensions to systematically assess argument quality. Given the abstract nature of ethos, pathos, and logos, automating their evaluation is complex. Moreover, while LLMs can produce fluent, persuasive arguments, these often lack credibility or sound logic (Kjeldsen, 2024). Thus, we aimed to establish objective criteria and stable evaluation methods for assessing rhetorical quality in GPT-generated content.

## 1.1 College Policy Debate

Competitive debate is a widely practiced extracurricular activity at both the high school and college levels, serving as an effective intervention for fostering critical thinking skills (Anderson & Mezuk, 2012). Few studies have explored the specific mechanisms involved or the potential of LLMs to support students in debate preparation. This study addresses that gap by focusing on evidence preparation in competitive debate, a process central to argument quality and participant success. Competitive debate in the United States consists of multiple formats, each following a structured format where debaters engage in rounds discussing a predetermined resolution. Participants, whether individuals or teams, adhere to set speaking times and are judged on the quality of their arguments. A key skill in debate is the persuasive presentation of high-quality evidence. This is demonstrated through the essential task of "cutting card", which involves the systematic collection and organization of evidence from sources such as news articles, blogs, peer-reviewed journals, and think tank reports. The "cutting card" process entails selecting relevant excerpts (referred to as **full text** thereafter), underlining key points (referred to as **key points** thereafter), highlighting sections for oral presentation (referred to as **highlights** thereafter), and crafting a "**tag**" as a concise, argumentatively framed summary of the evidence (Roush et al., 2024). Effective card-cutting requires advanced cognitive skills, including synthesis, emotional appeal, logical reasoning, and strategic framing, making it a high-level critical thinking exercise (Naqia et al., 2023). It also requires justifying selected arguments by logically linking premises to outcomes through accurate cause-and-effect or correlational reasoning. Hence, it is challenging for both human and AI debaters with a highly complex, multi-dimensional, and multi-step decision-making process.

## 1.2 Current Study

The current study explores GPT's ability to "cut cards", in which it is prompted to identify and extract effective arguments from full text. Specifically, we collected debate data consisting of the resolution and the cards that students created when preparing for a debate (Roush et al., 2024). Using the full text from a card, we prompted GPT to extract key points, highlights, and tags as if it were a student preparing for a debate. We analyzed the texts extracted by GPT and compared them to the texts selected and produced by students using a rubric developed based on Aristotle's rhetorical principles, which evaluate the persuasiveness of an argument based on three dimensions: ethos (credibility), pathos (emotional appeals), and logos (logical reasoning) (McCormack, 2014). We coded the three principles based on definitions and further developed eight highly associated indicators. This explorative study provides insights into the effectiveness of GPT in supporting debate preparation, highlighting its strengths and limitations in selecting persuasive arguments. Understanding how GPT's selections compare to those of human debaters can inform the development of AI-assisted debate tools and contribute to discussions on the role of AI in critical thinking and rhetoric.

## 2. Methods

### 2.1 Dataset

Our data is drawn from the OpenDebateEvidence dataset (Roush et al., 2024), a large-scale resource for argument mining and summarization, containing 3.5 million documents. As an explorative study, we randomly selected 30 evidence cards from the dataset generated by college students in Policy Debate (CX). The sample is stratified by side (16 affirmative and 14 negative cards) and is based on one resolution (topic). Debate evidence cards consist of several structured components, but our analysis focuses specifically on those that reflect argument selection and reasoning: the tag (a strategic summary), the citation (source attribution), underlined sections (key points), highlighted text (intended for oral delivery), and the length of each component. Following the same structure, we use ChatGPT-4-turbo to generate new evidence cards, integrating the tag, underlined, and highlighted text.

### 2.2 Prompt Engineering

The integration of large language models (LLMs) in qualitative research is reshaping data analysis methods (Barany et al., 2024; Xiao et al., 2023). In this study, we employed a zero-shot prompting approach to generate new evidence cards based on the same debate resolution and full text. Since students selectively extract quotations to support either the affirmative or negative stance, the LLM mirrors this process by using the same information and producing structured textual outputs aligned with debate learning. This method ensures systematic pattern identification while maintaining a structured analytical foundation. Notably, the initial prompt does not explicitly reference Aristotle's rhetorical principles or define ethos, pathos, and logos, but focuses solely on selecting arguments relevant to the debate. Since LLMs lack inherent knowledge of optimal spoken content duration for oral delivery, we integrated a standardized average length criterion for each variable in their outputs.

### 2.3 General Codebook for Aristotle's Rhetorical Principles

The development of this codebook followed an iterative process aimed at ensuring both conceptual clarity and practical applicability in evaluating debate performance among college students. Researchers first familiarize themselves with the dataset, identifying patterns and key themes related to ethos (credibility), pathos (emotional appeal), and logos (logical reasoning). Initial codes were established based on pre-defined theoretical constructs, while also considering the practical context of debate training. These preliminary codes were then refined through multiple cycles of application, revision, and discussion. Researchers applied the codes to raw data, assessed inter-rater agreement, and resolved discrepancies through collaborative discussions, ensuring a higher level of consistency and reliability. This iterative cycle continued until the coding framework reached a level of stability that effectively captured the nuances of debate performance. The resulting codebook provides a structured framework for evaluating the credibility of evidence, the emotional engagement of arguments, and the logical rigor of reasoning, with clear rating levels, definitions, and examples to guide analysis.

**2.4 Indicator Codebook for Aristotle's Rhetorical Principles**

To complement the qualitative coding process, we developed a series of quantitative indicators aligned with the themes of ethos, pathos, and logos. Prior work has attempted to fine-tune GPT models for detecting ethos, pathos, and logos in question-answer discussion forums (Moon et al., 2024), identifying various subdimensions of rhetorical appeals. While some of these indicators are not well-suited for debate preparation and argumentation, a refined set of indicators is presented, specifically designed for dismantling ethos, pathos, and logos. For example, the concepts of borrowed authority, factual knowledge, logical reasoning, and statistics are adapted from Moon et al.'s codebook. Ethos is assessed through (1) borrowed expert authority, (2) borrowed organizational authority, and (3) literature credibility. Pathos is identified through (1) threat amplification and (2) empathic resonance. Logos is evaluated based on (1) factual knowledge, (2) logical reasoning, and (3) statistical evidence. Each indicator is further defined and illustrated with examples and results of quantitative measurement in Table 2. To clarify this process, an easy-to-understand workflow is illustrated in Figure 1.

Figure 1. Debate Performance: Student vs. GPT-4o Rhetorical Analysis

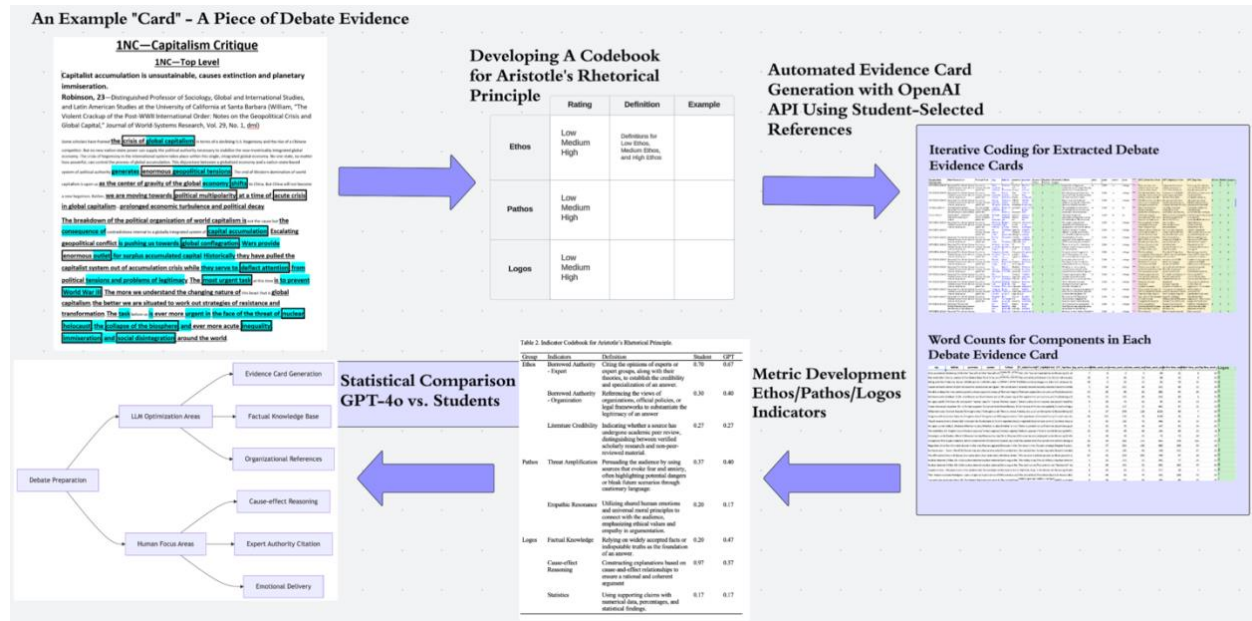

## 3. Results

Based on our qualitative analysis of ethos, pathos, and logos, we consulted a debate expert with over 10 years of experience, and together we interpreted the performance of both the student and GPT-4o. Analysis of full-text quotations showed that student-generated tags averaged 15.7 words, while GPT averaged 70.7 words. Students' summaries averaged 138.4 words compared to GPT's 70.7, and their spoken content averaged 67.4 words versus GPT's 24.3. The word count reflects the amount of linguistic content used for persuasion and evidence presentation. If tags, summaries, and spoken content serve similar purposes in this context, then using fewer words to achieve the same rhetorical goal may indicate lower linguistic effort—signaling greater persuasive efficiency. We measured the ethos, pathos, and logos of debate preparation processes of students and GPT-4o. There is no significant difference in the overall scoring of ethos (credibility) between college students and GPT-4o. This outcome was anticipated, as both groups utilized identical foundational literature for standardized high-level comparisons. A Mann-Whitney U test revealed no significant difference in pathos (emotional appeals), $z = -0.33$, $p = 0.741$, $n = 30$. However, logos (logical reasoning) differed significantly, $z = 4.61$, $p < 0.001$, $n = 30$. These findings suggest that while college students and GPT-4o perform similarly in ethos and pathos, students significantly outperform GPT-4o in logical reasoning in selecting effective arguments. Regarding subdimension indicator detection, Table 2 presents the probability of occurrence for each indicator. The probabilities of these indicators appearing in student-generated and LLM-generated evidence cards show no significant differences for most ethos and pathos indicators. However, notable differences emerge in the domain of logos. Specifically, GPT-4o–generated content contains more factual knowledge, whereas student-generated content demonstrates a higher frequency of cause-and-effect reasoning.

## 4. Discussion and Conclusion

The most notable difference between student and GPT performance lies in logical reasoning, particularly in constructing cause-and-effect relationships. A Mann–Whitney U test showed that students were significantly more likely to detect cause-and-effect reasoning between arguments than GPT, resulting in higher logos scores. Our professional debate trainer also noted a critical factor in live debates that GPT's spoken delivery is less efficient than that of students, which can be measured by word economy per effective message. Beyond these metrics, results also capture performance on meta-level tasks such as winning rounds, adhering to time limits, and persuading judges. GPT generates concise summaries and spoken content but produces longer, less efficient tags, indicating a structural preference for brevity in key point extraction while overextending in tagging. Students' superior logical reasoning highlights the effectiveness of human argument construction. Future research will refine human-AI comparisons by analyzing factors like training data, prompt structure, and response generation. These subdimension indicators can aid in the automated detection of rhetorical patterns in persuasion and argumentation. Our new framework integrates thematic analysis and detailed indicators to systematically assess the quality of arguments. Additionally, efforts will focus on enhancing inter-rater reliability in qualitative assessments and expanding datasets beyond thirty cases.

# References

Anderson, S., Mezuk, B.: Participating in a policy debate program and academic achievement among at-risk adolescents in an urban public school district: 1997–2007. Journal of Adolescence. 35(5), 1225–1235 (2012).

Barany, A., Nasiar, N., Porter, C., Zambrano, A. F., Andres, A. L., Bright, D., Shah, M., Liu, X., Gao, S., Zhang, J., Mehta, S., Choi, J., Giordano, C., Baker, R. S.: ChatGPT for Education Research: Exploring the Potential of Large Language Models for Qualitative Codebook Development. In: Proceedings of the 25th International Conference on Artificial Intelligence in Education, pp. 134-139. Springer, Heidelberg (2024).

Braet, A. C.: Ethos, pathos and logos in Aristotle's Rhetoric: A re-examination. Argumentation, 6, 307-320 (1992).

Kjeldsen, J. E.: Ethos, Technology, and AI in Contemporary Society. 1st edn. Routledge, London (2024)

McCormack, K. C.: Ethos, pathos, and logos: The benefits of Aristotelian rhetoric in the courtroom. Washington University Jurisprudence Review 7 (1), 131-155 (2014).

Moon, H., Bae, B. J., & Bae, S.: Developing a ChatGPT-Based Text Extraction Model to Analyze Effective Communication Elements in Pandemic-Related Social Q&A Responses. In: 2024 International Conference on Artificial Intelligence in Information and Communication (ICAIIC), pp. 728–731. IEEE, Osaka, Japan (2024).

Naqia, D., As' ari, A. A., & Suaidi, A. (2023). Students' critical thinking skills perform in debate activities. Journal of English Language Teaching and Cultural Studies, 6(1), 55-64.

Roush, A., Shabazz, Y., Balaji, A., Zhang, P., Mezza, S., Zhang, M., et al.: Opendebateevidence: A massive-scale argument mining and summarization dataset. Preprint at http://arxiv.org/abs/2406.14657 (2024).

Schueler, B. E., & Larned, K. E.: Interscholastic Policy Debate Promotes Critical Thinking and College-Going: Evidence from Boston Public Schools. Educational Evaluation and Policy Analysis 43(3), (2023).

Xiao, Z., Yuan, X., Liao, Q. V., Abdelghani, R., Oudeyer, P. Y.: Supporting qualitative analysis with large language models: Combining codebook with GPT-3 for deductive coding. In: Companion Proceedings of the 28th International Conference on Intelligent User Interfaces, pp. 75–78. Association for Computing Machinery, New York (2023).

# Appendices

Table 1. General Codebook for Aristotle's Rhetorical Principles.

| Group | Rating (0-2) | Definition | Examples |
|---|---|---|---|
| Ethos (credibility) | Low | The evidence resource is not reliable and cannot be found based on citation. | "Quora contributors. (n.d.). Is it time to begin formal political debates on the rights of sentient robots. Quora. Retrieved February 3, 2025" |
| | Medium | While the evidence source is identifiable, articles from personal websites, discussion forums, or less reliable news outlets are not rated as 2 unless they are peer-reviewed publications. | "He is also a trained United Nations Missions Observer and participated in several EU and NATO missions.], 2018, "Autonomous Weapon Systems in International Humanitarian Law," Joint Air Power Competence Centre" |
| | High | The source is identifiable, and the article is a peer-reviewed publication from conferences or journals. | "Smith, P. T. (2019). Just research into killer robots. Ethics and information technology, 21, 281-293." |
| Pathos (emotional appeals) | Low | The content is primarily factual and objective, with neutral, detached language and no attempt to evoke emotional responses. Emotionally charged words or anecdotes are avoided. | "Justice as a moral concept means the impartial application of rules of conduct are impartial. A just world is a world in which distributive considerations are subordinated to moral ones not the reverse" |
| | Medium | The content includes elements that elicit some emotional response, using descriptive language and relatable examples to connect with the audience on a personal level. | "The value of necessities deemed essential by society make up the value of labor power. Any place where bosses can hold down wages ensures a cheaper labor pool among the oppressed demographic, but also that everyone's wages are pushed down." |
| | High | The content strongly evokes emotions through vivid imagery, powerful anecdotes, and emotionally charged language. | "…human soldiers often fail to discriminate or engage in disproportionate uses of force because of anger, fatigue, or fear; LAWS experience none of these. The use of LAWS can have a spiraling positive effect on these dynamics where fewer warfighters lead to…" |
| Logos (logical reasoning) | Low | The summary is inaccurate, the tag contradicts the authors' perspectives, and the evidence fails to support the resolution effectively. | "…good is always an attributive adjective Something may be a good, but nothing is good period emotivism and related accounts of the meaning of good cannot be generally correct there is no such property as being a good pebble, good act, and so on." |
| | Medium | The summary accurately represents the original text and strengthens the resolution's premises, though rhetorical strategies may involve logical fallacies or exaggerations of the evidence. | "…not sufficient evidence that the Japanese public wants; not enough resources budgeted for defense." |
| | High | The tag and spoken summary are precise, logically sound, and free of fallacies. The evidence aligns perfectly with the resolution's purpose. | "…if Turkish confidence in the U.S. and NATO commitment to its security weakens, Ankara could begin to explore other options, including nuclear deterrence." |

Table 2. Indicator Codebook for Aristotle's Rhetorical Principle.

| Group | Indicators | Definition | Student | GPT-4o |
|---|---|---|---|---|
| Ethos | Borrowed Authority - Expert | Citing the opinions of experts or expert groups, along with their theories, to establish the credibility and specialization of an answer. | 0.70 | 0.67 |
| | Borrowed Authority - Organization | Referencing the views of organizations, official policies, or legal frameworks to substantiate the legitimacy of an answer | 0.30 | 0.40 |
| | Literature Credibility | Indicating whether a source has undergone academic peer review, distinguishing between verified scholarly research and non-peer-reviewed material. | 0.27 | 0.27 |
| Pathos | Threat Amplification | Persuading the audience by using sources that evoke fear and anxiety, often highlighting potential dangers or bleak future scenarios through cautionary language. | 0.37 | 0.40 |
| | Empathic Resonance | Utilizing shared human emotions and universal moral principles to connect with the audience, emphasizing ethical values and empathy in argumentation. | 0.20 | 0.17 |
| Logos | Factual Knowledge | Relying on widely accepted facts or indisputable truths as the foundation of an answer. | 0.20 | 0.47 |
| | Cause-effect Reasoning | Constructing explanations based on cause-and-effect relationships to ensure a rational and coherent argument | 0.97 | 0.37 |
| | Statistics | Using supporting claims with numerical data, percentages, and statistical findings. | 0.17 | 0.17 |